\documentclass{elsart}

\usepackage{graphicx}
\usepackage[latin2]{inputenc}
\usepackage[T1]{fontenc}

\usepackage{amssymb}

\begin{document}

\begin{frontmatter}

\title{The influence of the time delay of information flow on an economy evolution. \\
  The stock market analysis.}
\author{Janusz Miśkiewicz}
\ead{jamis@ift.uni.wroc.pl}
\address{Institute of Theoretical Physics, Wroc\l{}aw University \\ pl. M. Borna 9, 50-204 Wroc\l{}aw, Poland}

\begin{abstract}
The decision process requires information about the present state of the system, but in economy acquiring data and processing them is an expensive and time consuming process. Therefore the state of the system is measured and announced at the end of the well defined time intervals.
The model of a stock market coupled with an economy is investigated and the role of the length of the time delay of information flow investigated. It is shown that increasing the time delay leads to collective behavior of agents and oscillations of autocorrelations in absolute log-returns. 
\end{abstract}

\end{frontmatter}

\section{Time delay in economy systems}
\label{sec:time_del}
The most important part of contemporary management board work is processing informations. It means that based on various informations e.g. exchange rate, demand and request level, stock market data, inflation, productivity, unemployment, householder income, country budget, government and central bank decisions etc. they have to make plans and decisions. On the other hand the ''measurement'' process in economy is an expensive and time consuming task. Considering the size of any economy system one has to admit that simple collecting data require time, what more processing the raw data is a tremendous work e.g. GDP of Poland for the year 2006 was announced in June 2007. So the time delay between the actual state of the system and the moment when the information is known is significant. Moreover the ''measurement'' process is expensive. Therefore the state of the system is not observed continuously but checked at discrete and well defined time intervals. The time delay resulting from the above reasons is specific to the data and market. It can vary from seconds for a stock exchange markets up to nearly a year for some agricultural markets.

The relationship between stock market and economy is well known and discussed in newspapers and specialized magazines. The special attention is usually payed to the government and central bank decisions and theirs influence onto the behavior of the stock market. For example the cut of interest rate by US Federal Reserve resulted in significant increase of market indexes e.g. S\&P 500 increased 1.59\% within 15min after the announcement, the main Wall Street indexes increased by 2.51\% - 2.91\% at the end of the trading day. So there is no need to discuss the obvious fact that stock market takes into account external informations. Of course in real situation many different factors influence simultaneously the stock market and it is not possible to separate only one factor in order to investigate its impact. Therefore modelling the stock market \cite{stock_market1,stock_market2,stock_market3,stock_market4} gives the only possibility to observe features which are usually hidden among many others. 

The influence of the time delay on the behavior of economy system was investigated in previous works \cite{JMMA1,JMMA2}. In this studies the ACP model of interacting agents was. The model was originally set by M. Ausloos, P. Clippe and A. Pękalski in order to investigate economy systems under changing political conditions \cite{ACP4,ACP1,ACP2,ACP3}. The ACP model was then modified \cite{JMMA1,JMMA2} in order to split the information about the system into two parameters: the real state of the system and the information about the system which is updated in well defined time intervals. As far two cases have been investigated: the homogeneous \cite{JMMA1} and heterogeneous \cite{JMMA2} concentration of companies.
It has been shown that in the case of homogeneous concentration of companies the extension of the time delay leads to economy cycles and even to crashes of the system. On the other hand in heterogeneous systems we do not observe crashes of the whole system (they happen only locally) but the delay time of information flow is responsible for economy cycles.

Within this paper, in difference from the previous work where the general model of the economy was considered, the influence of time delay of information flow is considered in the case of stock market model. The special attention is payed to the stock market because it is considered as a system with instantaneous information flow, but in fact it is coupled to an economy system, in which the time delay of information flow can not be neglected.

\section{Model description}
\label{sec:model}
In order to investigate the role of the time delay in information flow the Ising model of financial markets \cite{sornette} was modified by introducing interaction between the stock market and the economy as well as the time delay in the information transfer about the economy state. The main assumptions of the model are:

\begin{itemize}
\item on the stock market there are ${\cal N}$ interacting agents;
\item at every trading events they have to take one of the two possible decisions: sell or buy;
\item the decision depends upon:
\begin{itemize}
\item decisions of other agents (the friendship matrix $K_{ij}(t)$),
\item public informations about the economy state ({\it is updated at discrete time intervals}),
\item private informations about market state;
\end{itemize}
\item the market state is considered random. 
\end{itemize}
\subsection{Model definition}
The decision matrix:
\begin{equation}
\label{eq:decyzja}
s_i(t) = sign \left[ \sum_{j \in {\cal N}} K_{ij}(t) s_j (t-1) + \sigma_i (t) G(t) + \epsilon_i (t) \right], 
\end{equation}
where
$\sigma_i (t)$ -- sensitivity to the external news (the state of the economy), $G(t)$ -- external news, normally distributed and updated at the time $t=n \cdot t_d$, $n\in (1,2,3,\ldots)$, $\epsilon_i (t)$ --  private news defined as a uniform random variable from the interval $[CV, CV+0.1]$. \newline
The friendship matrix:
\begin{equation}
\label{eq:friendship_matrix}
K_{ij}(t) = b_{ij} + \alpha K_{ij}(t-1) + \beta r(t-1)G(t-1), 
\end{equation}
where
$b_{ij} $ -- initial trust matrix, values are uniformly distributed in the interval $ (0,b_{max})$, $r(t)$ -- return of the market, $\alpha, \beta $ -- model parameters. \newline
The return and price:
\begin{equation}
\label{eq:return}
r(t)=\frac{\sum_{i \in {\cal N}} s_i(t)}{\lambda N}, \;\;\;\; p(t) = p(t-1) \exp (r(t)),
\end{equation}
$\lambda$ -- market depth.\newline
The economy state is assumed to be random and is sampled from the normal distribution:  $G(t)=0.1 \cdot N(0,1)$. The simulations were performed up to $10^5$ iteration steps (IS), which resulted in time series of $10^5$ data points.

\begin{figure}
 \centering
 \includegraphics[bb=50 50 410 302]{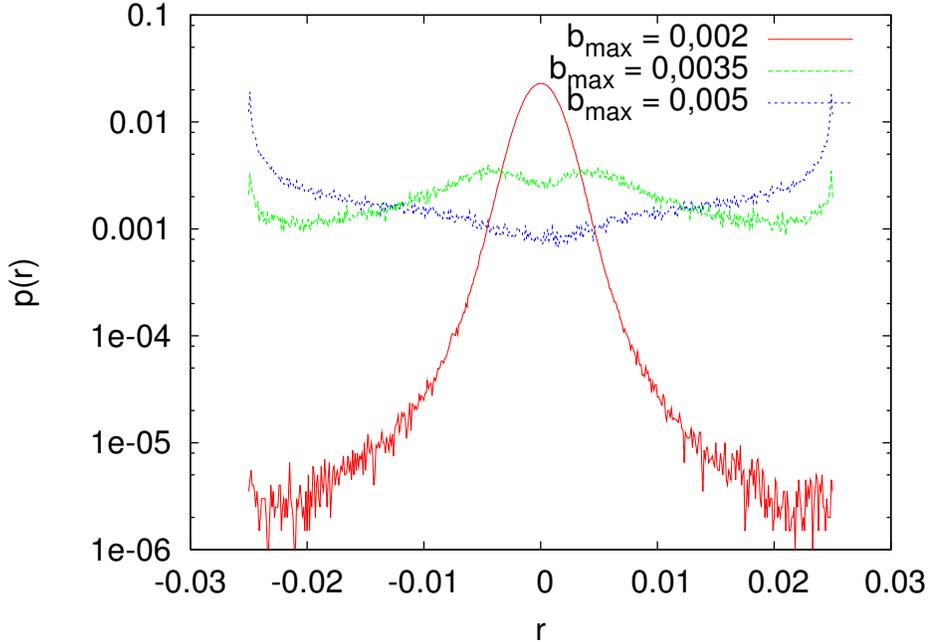}
 \caption{(Color online) Probability distribution of stock market returns in semilog plot. The stock market model coupled with random economy. No time delay in information flow. The model parameters: $\alpha = 0.1$, $\beta=1$, $CV=0.2$, $\sigma_{max}=0.04$, $\lambda= 40$, ${\cal N}=500$, $IS=10^5$}
 \label{fig:model_pods}
\end{figure}

The detailed analysis of the applied stock market model can be found in \cite{sornette}. For the sake of clarity the most important features will be recalled. The decisions of an agent (Eq.(\ref{eq:decyzja})) depends on the market state and opinions of other agents according to the friendship matrix Eq.(\ref{eq:friendship_matrix}). The friendship matrix is not assumed symmetric. On the other hand the $b_{max}$ parameter decides upon the ''strength'' of the friendship between agents. Considering the case without interaction between stock market and economy one can observe that for high values of friendship parameter e.g. $b_{max} > 0.003$ for the stock market model defined by:  
$\alpha = 0.1$, $\beta=1$, $CV=0.2$, $\sigma_{max}=0.04$, $\lambda= 40$, ${\cal N}=500$, $IS=10^5$ the highest values of stock market return probability distribution (Fig.\ref{fig:model_pods}) are for $r=-0.025$ or $r=0.025$, which are the most extreme return values possible for this particular system. This means that increase of $b_{max}$ parameter leads to the collective behavior of agents.

\section{Results}
\label{sec:results}

\begin{figure}
 \centering
 \includegraphics[bb=50 50 410 302]{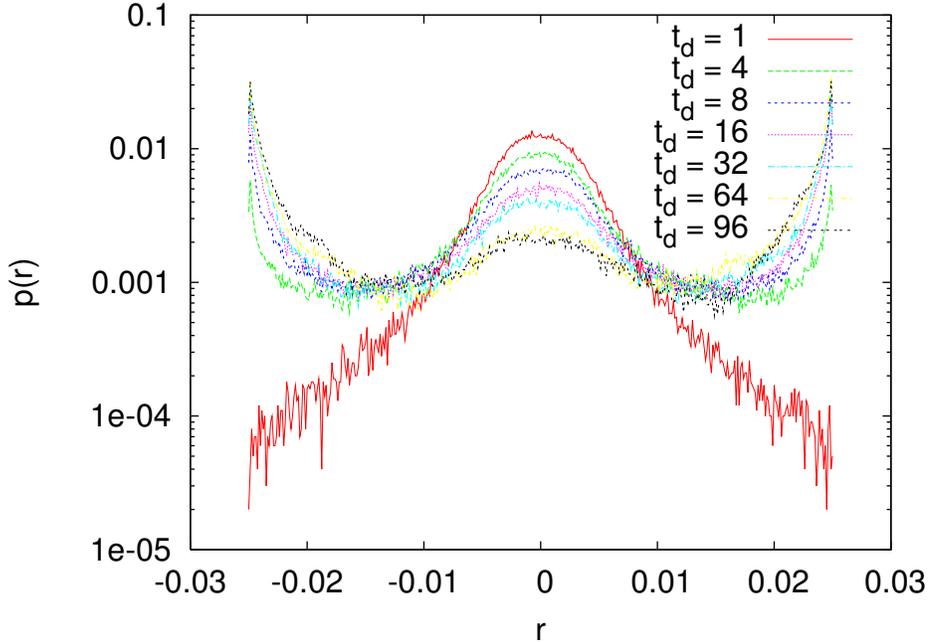}
 \caption{(Color online) Probability distribution of stock market distributions in semilog plot. The stock market model coupled with the random economy, $t_d \geq 1 \; IS$. Model parameters: $\alpha = 0.1$, $\beta=1$, $CV=0.2$, $\sigma_{max}=0.04$, $b_{max}=0.0033$, $\lambda= 40$, ${\cal N}=400$, $IS=10^5$. }
 \label{fig:return_del}
\end{figure}

In order to investigate the influence of time delay in information flow between the economy and the stock market the following stock market model was used: $\alpha = 0.1$, $\beta=1$, $CV=0.2$, $\sigma_{max}=0.04$, $b_{max}=0.0033$, $\lambda= 40$, ${\cal N}=400$, $IS=10^5$ and analyzed with different values of time delay parameter $t_d = 1,4,8,16,32,64,96$ $ IS$. The probability distributions function (PDF) of stock market returns are presented in Fig.\ref{fig:return_del}. In the case of the lowest value of the time delay parameter $t_d=1  \; IS$ maximum of PDF is at its mean $r=0$ and it is similar to the case with instantaneous transfer of informations between the stock market and the economy (Fig.\ref{fig:model_pods}). But even small increase of the time delay ($t_d=4 \; IS$) results in strong increase of the collective behavior of agents. The probability of all agents taking the same decision is approximately 500 times higher than at the case of $t_d= 1 \; IS$. The form of PDF is also changing. For the system without time delay or $t_d=1 \; IS$ PDF has one maximum at $r=0$ and besides this point PDF is monotonic, while for $t_d \geq 4 \; IS$ PDF has three maximums and two minimums. Moreover the extension of the time delay results in decreasing of the maximum at point $r=0$ and increasing probability density at the remaining two maximums at $r=-0.025$, $ r=0.025$.

\begin{figure}
 \centering
 \includegraphics[bb=50 50 482 302]{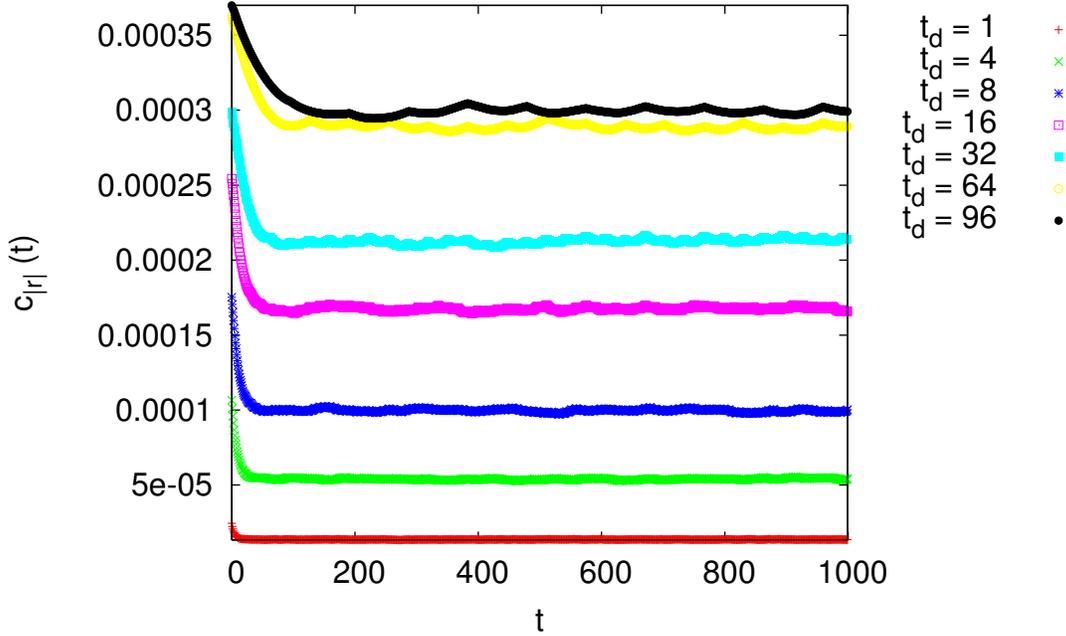}
 \caption{(Color online) The autocorrelation function of absolute log return for a chosen time delay of information flow}
 \label{fig:corr_zwrot}
\end{figure}

\begin{figure}
 \centering
 \includegraphics[bb=50 50 410 302]{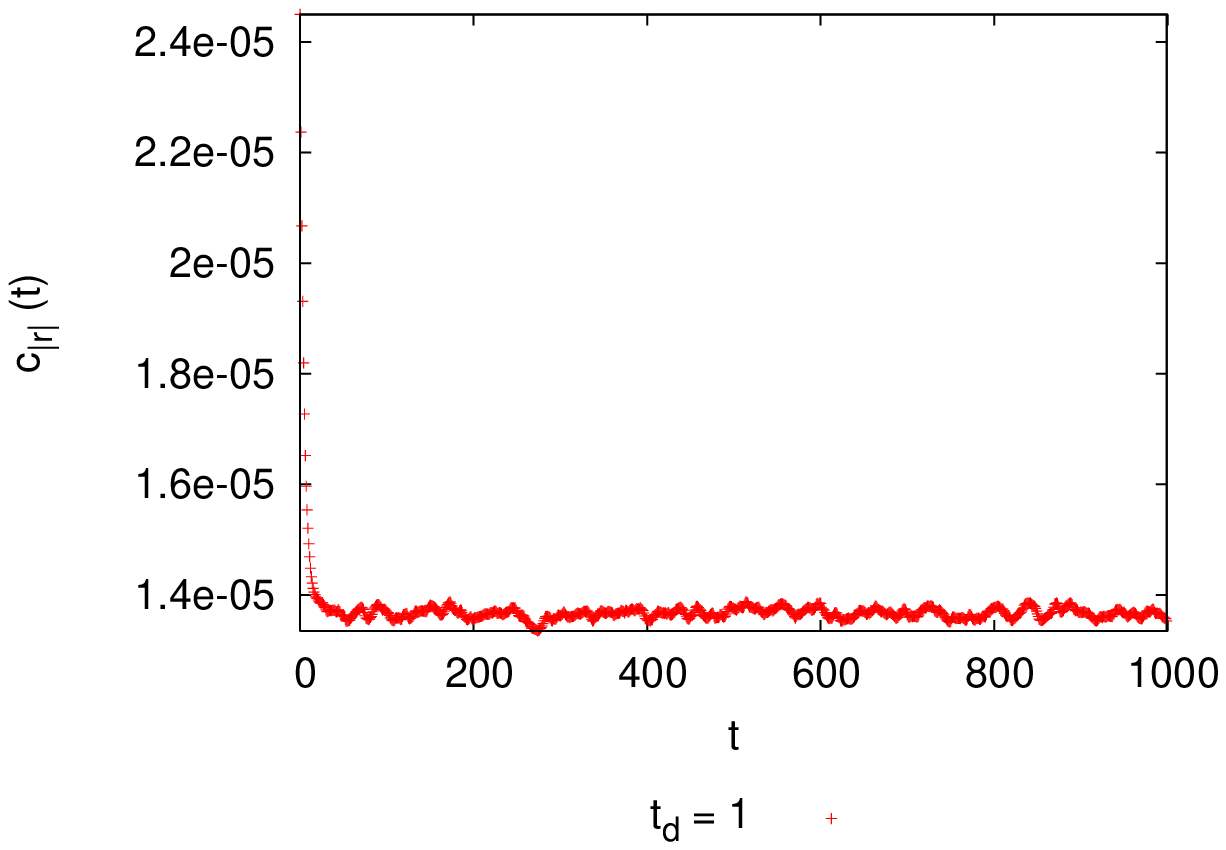}
 \caption{The autocorrelation function of absolute log return $t_d=1$}
 \label{fig:corr_zwrot_t1}
\end{figure}

\begin{figure}
 \centering
 \includegraphics[bb=50 50 410 302]{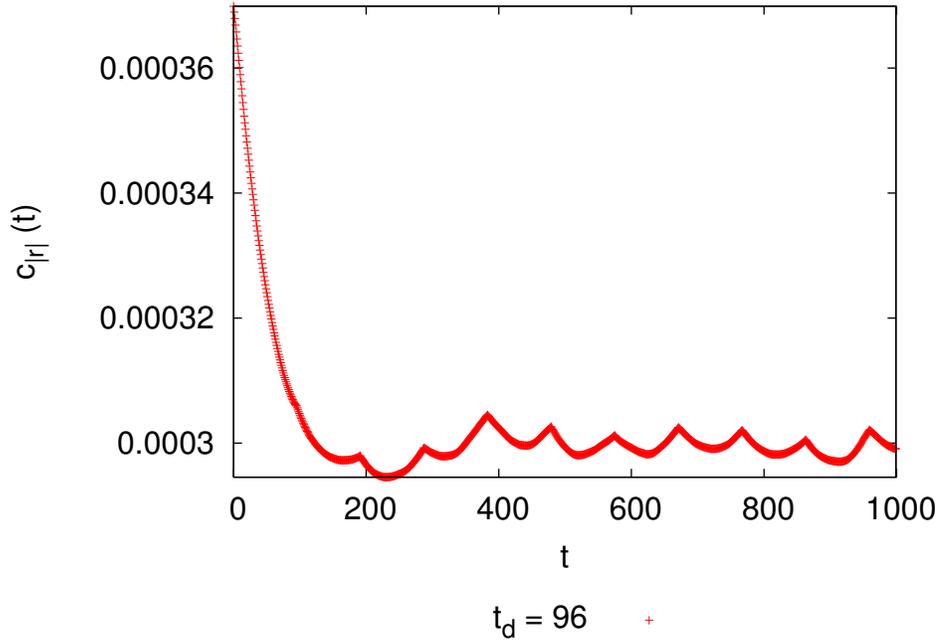}
 \caption{The autocorrelation function of absolute log return $t_d=96$}
 \label{fig:corr_zwroty_t96}
\end{figure}

Since the PDF function of the system with time delay of information flow showed existence of collective behavior of the agents the autocorrelation function of absolute log-returns given by Eq.(\ref{eq:zwroty}) was calculated. 
\begin{equation}
\label{eq:zwroty}
 c_{|r|} (\tau) = \langle | r(t+\tau ) | |r(t)| \rangle,
\end{equation}
where the brackets $\langle ... \rangle$ denotes calculating the mean values.
The autocorrelation of absolute log-returns as a function of time lag for different values of $t_d$ are presented in Fig.\ref{fig:corr_zwrot} and for the sake of clarity two special cases: $t_d=1 \; IS$ and $t_d=96 \; IS$ are presented separately Fig.\ref{fig:corr_zwrot_t1} and \ref{fig:corr_zwroty_t96} respectively. The mean value of $c_{|r|} $ for chosen time delay of information flow are presented in Tab.\ref{tab:korelacje}. It is varying from $1.4 \cdot 10^{-5}$ for $t_d=1 \; IS$ up to $3 \cdot 10^{-4}$ for $t_d = 96 \; IS$. Since the highest value is 21 times greater than the lowest value the difference is considered as significant. Comparing the autocorrelation of absolute log-returns for the case of $t_d=1 \; IS $ (Fig.\ref{fig:corr_zwrot_t1}) with $t_d=96 \; IS$ (Fig.\ref{fig:corr_zwroty_t96}) it can be noticed that increasing $t_d$ results in periodic behavior of the autocorrelation function of absolute log-returns with the period of $86 \; IS$ approximately.
\begin{table}
\begin{tabular}{|l|c|c|c|c|c|c|c|}
\hline
 $t_d$ & 1 & 4 & 8 & 16 & 32 & 64 & 96 \\ \hline
$\langle c_{|r|}\rangle $ & $1.4 \cdot 10^{-5}$ & $5.5 \cdot 10^{-5}$ & $ 10^{-4}$ & $1.66 \cdot 10^{-4}$ & $2.15 \cdot 10^{-4}$ & $2.9 \cdot 10^{-4}$ & $3 \cdot 10^{-4}$ \\ \hline
\end{tabular} 
\label{tab:korelacje}
\caption{The mean values of the autocorrelation function of absolute log-returns for the given time delay of information flow.}
\end{table}

\begin{figure}
 \centering
 \includegraphics[bb=50 50 410 302]{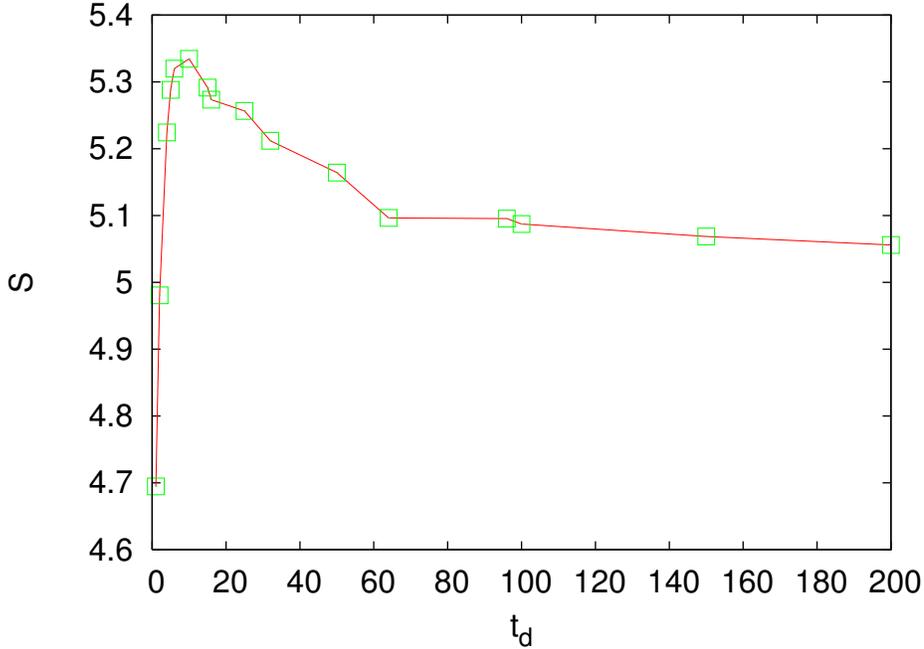}
 \caption{Shannon information entropy of the return distribution}
 \label{fig:entropy}
\end{figure}

The Shannon information entropy of the return distribution defined by Eq.(\ref{eq:shanon}) as a function of the time delay length is presented in Fig.\ref{fig:entropy}. 
\begin{equation}
\label{eq:shanon}
 S(t_d)=-\sum p_i \log_2 p_i
\end{equation}
The Shannon entropy has maximum at $t_d= 10 \; IS$ then is decreasing relatively quickly with the slope $-4.4 \cdot 10^{-3}$ up to the time delay $t_d=64 \; IS $ to become almost constant -- the fitted linear function has the coefficient approximately ten times smaller (slope coefficient $ \approx -3 \cdot 10^{-4}$). This observation is with agreement with the analysis of autocorrelations of absolute log-returns. The difference between $c_{|r=96 \; IS|} $ and $c_{|r= 64 \; IS|} $ is significantly smaller than between any other values of $c_{|r|} $. This suggest that the system achieves a saturation like state. Even for significantly longer values of $t_d$ the behavior of agents do not change significantly -- they are highly unanimous in theirs decisions, the autocorrelation function of the absolute returns do not change much as well as the complexity of stock market returns measured by Shannon information entropy.

\section{Conclusions}
The interaction between economy and stock market is a fascinating and relevant
subject of interest in many economy questions \cite{economy_interr1,economy_interr2}.
The problem is studied by investigating stochastic properties of stock markets returns \cite{s1,s2,s3,s4,s5} or correlation between different economies \cite{gdp1}.
In the present analysis, an information flow, is incorporated into the stock market model interacting with external field -- economy. This has led to observe for relative long time delay ($t_d \geq 32 \; IS$) so called cycles, through autocorrelation function of absolute returns. The time delay in information flow results in stimulating collective behavior of agents and this feature is seen even for short time delay ($t_d \geq 4 \; IS$). The existence of resonance time delay, for which the entropy of returns values achieves its maximum seams to be an interesting result of the present study since in this case the forecasting is the most difficult. Moreover the second specific time delay can be distinguished: the critical delay time, above which the system features do not change significantly -- for the model investigated here it is $t_d=64 \; IS$.
The later observations lead us to the conclusion that the analysis of the time delay of information flow may help us to better understand stock market processes and control its behavior.


\bibliographystyle{elsart-num} 
\bibliography{apfa_6}
\end{document}